\documentclass[letterpaper, 10 pt, conference]{ieeeconf}  

\IEEEoverridecommandlockouts                              

\usepackage{graphicx}
\usepackage{amsmath}
\usepackage{amssymb}
\usepackage{amsthm}
\usepackage[english]{babel}
\usepackage[utf8]{inputenc}
\usepackage{algorithm}
\usepackage[noend]{algpseudocode}
\usepackage{cleveref}
\usepackage{times} 
\usepackage{mathtools}
\usepackage{booktabs}
\usepackage{bm}
\usepackage{wrapfig}
\usepackage{xcolor}
\usepackage{tikz}
\usepackage{pgfplots}
\pgfplotsset{compat=newest}
\usepackage{subcaption}
\usepackage[font={footnotesize}]{caption}

\newcommand{\bp}{\bm{p}}

\newcommand{\bx}{\bm{x}}

\newcommand{\bI}{\mathbf{I}}

\newcommand{\cJ}{\mathcal{J}}

\newcommand{\cU}{\mathcal{U}}

\newcommand{\cX}{\mathcal{X}}

\newcommand{\II}{\mathbb{I}}

\newcommand{\RR}{\mathbb{R}}

\newcommand{\argmin}{\mathop{\mathbf{argmin}}}

\newcommand{\minimize}{\mathop{\mathbf{min}}}

\newcommand{\minimizewrt}[1]{\mathop{\underset{#1}{\minimize}}}

\newcommand{\argminwrt}[1]{\mathop{\underset{#1}{\argmin}}}

\newcommand{\defeq}{\vcentcolon=}

\newcommand{\st}{\mathop{\mathbf{s.t.}}}

\newtheorem{problem}{Problem}

\newtheorem{remark}{Remark}[problem]
\newtheorem*{remark*}{Remark}

\colorlet{revision_color}{black}
\newcommand{\revcolor}[1]{\textcolor{revision_color}{#1}}

\newcommand{\ie}{\textit{i}.\textit{e}., }
\newcommand{\eg}{\textit{e}.\textit{g}., }

\title{\LARGE \bf
Real-time Safety Index Adaptation for Parameter-varying Systems\\via Determinant Gradient Ascend
}

\author{Rui Chen$^{1}$, Weiye Zhao$^{1}$, Ruixuan Liu$^{1}$, Weiyang Zhang$^{2}$, and Changliu Liu$^{1}$
\thanks{$^{1}$Carnegie Mellon University, Pittsburgh, PA. Contact: {\tt\small \{ruic3, weiyezha, ruixuanl, cliu6\}@andrew.cmu.edu}}%
\thanks{$^{2}$University of Michigan, Ann Arbor, MI. Contact: {\tt\small zhangwy@umich.edu}}%
\thanks{This work is partially supported by the National Science Foundation, Grant No. 2144489.}%
}

\begin{document}

\maketitle
\thispagestyle{empty}
\pagestyle{empty}

\begin{abstract}
Safety Index Synthesis (SIS) is critical for deriving safe control laws.
Recent works propose to synthesize a safety index (SI) via nonlinear programming and derive a safe control law such that the system 1) achieves forward invariant (FI) with some safe set and 2) guarantees finite time convergence (FTC) to that safe set.
However, real-world system dynamics can vary during run-time, making the control law infeasible and invalidating the initial SI.
Since the full SIS nonlinear programming is computationally expensive, it is infeasible to re-synthesize the SI each time the dynamics are perturbed.
To address that, this paper proposes an efficient approach to adapting the SI to varying system dynamics and maintaining the feasibility of the safe control law.
The proposed method leverages determinant gradient ascend and derives a closed-form update to safety index parameters, enabling real-time adaptation performance.
A numerical study validates the effectiveness of our approach.

\end{abstract}

\section{Introduction}\label{sec:intro}




Autonomous systems are entering many application domains, \eg autonomous vehicles \cite{9046805}, human-robot collaboration \cite{9500153, liu2022safe}, etc.
As autonomous systems are deployed to more dynamic environments, safety becomes increasingly critical. 
It is important to ensure that the system would not harm the agents sharing the environment (\ie humans and the workspace).

Safe control has been widely studied to guarantee the safety of autonomous systems.
In particular, energy functions \cite{wei2019unified} are widely used in the safe control field to quantify system safety and derive control laws to ensure safety, such as the safe set algorithm (SSA) \cite{liu2014control} and control barrier functions (CBF) \cite{ames2014control}.
To achieve provable safety, the safe control law needs to satisfy two critical properties: 1) \textbf{\textit{forward-invariance}} (FI), meaning that the system should stay in a safe region once entering it, and (b) \textbf{\textit{finite-time convergence}} (FTC), meaning that the system should land in the safe region in finite time even starting in an unsafe state.
To achieve such a provably safe control law, a safety index (SI) needs to be carefully synthesized so that the constraints yield from the SI is always feasible.
Namely, in every state of interest, there must exist a control in the control space (either bounded or unbounded), that satisfies the safety constraints.
Therefore, \textit{Safety Index Synthesis} (SIS) is critical \cite{zhao2021zeroviolation, zhao2023sos}.

\begin{figure}[t]
    \vspace{10pt}
    \centering
    \includegraphics[width=\linewidth]{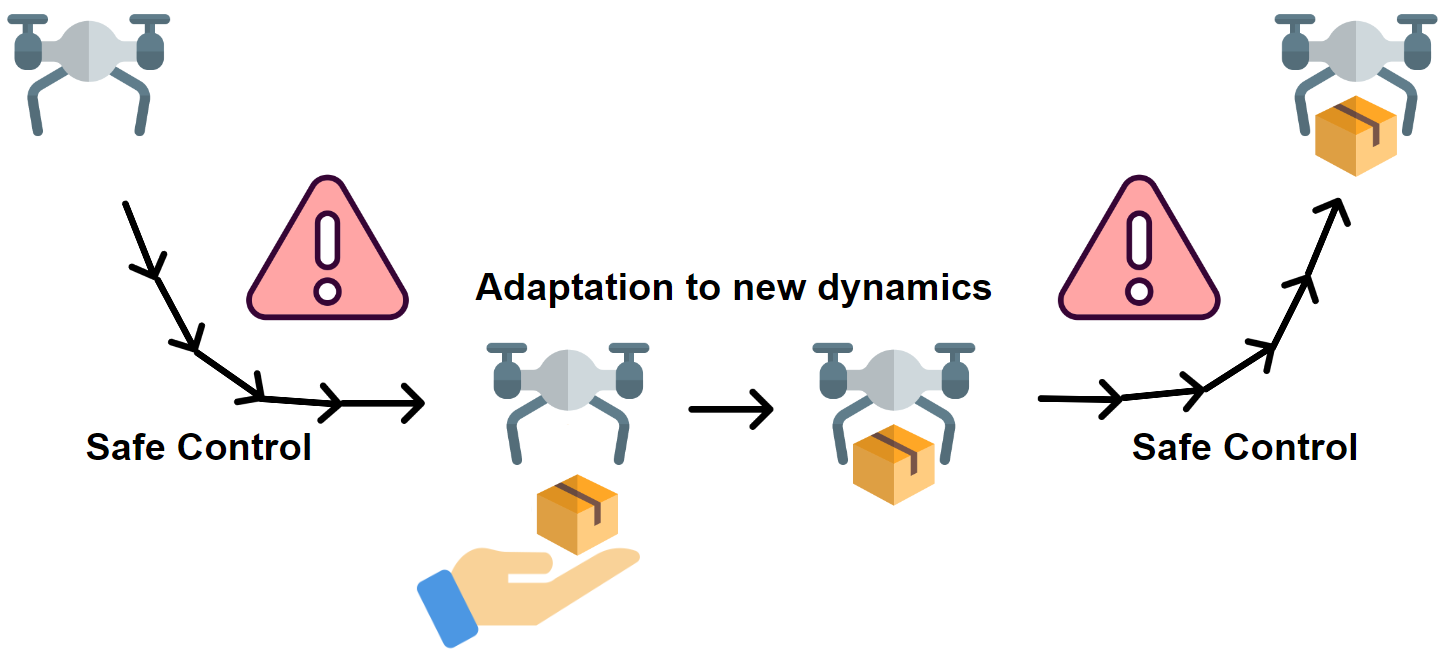}
    \caption{Illustration of safety index adaptation. After the drone picks up a package whose weight is not known in advance, its dynamics change. The safe control law is adapted to the new dynamics and continues to guarantee safety, e.g., collision avoidance.}
    \label{fig:intro}
    \vspace{-15pt}
\end{figure}

SIS has been widely studied.
Previous works \cite{ma2022joint,dawson2022safe} address SIS for dynamic systems with unbounded control.
\cite{zhao2021zeroviolation,wei2022nndm,zhao2023sos} address SIS for systems with known bounded control.
Recent work \cite{chen2023sis} further addresses the SIS problem for dynamic systems with varying (\ie state-dependant) control bounds, which is more practical in reality.
Although existing approaches are promising, most of them consider invariant dynamic systems.
In practice, the dynamics of real-world systems are usually varying.
For example, when a drone is used for package delivery, its dynamics change every time a package is added or removed (see \Cref{fig:intro}); when a robot arm is used for pick-and-place, its dynamics can change due to the object being manipulated.
Under perturbed dynamics, the safe control law derived from the previous safety index might no longer be feasible, and can no longer guarantee safety.
A naive fix is to re-synthesize the SI whenever the dynamics change.
However, a full SIS generally requires non-trivial efforts and is infeasible for real-time adaptation.
For instance, it can take more than 10 minutes to synthesize a single SI for a simplistic unicycle model with state-dependant control bounds \cite{chen2023sis}.

This paper studies efficient \textit{safety index adaptation} (SIA) for parameter-varying systems.
Our intuition is that when the system dynamics change, it should be sufficient to fine-tune the safety index instead of generating a new one from scratch.
To achieve that, we first observe that the full SIS problem is in fact solved via a semidefinite program with a positive-semidefiniteness (PSD) constraint that depends on the system dynamics.
That constraint is normally violated when the dynamics change, invalidating the previous safety index.
A reasonable solution is to fine-tune the SI parameters such that the PSD constraint is satisfied again.
Leveraging Sylvester's criterion \cite{horn2012matrix}, we are able to derive closed-form updates to the SI parameters that are computationally efficient enough for real-time adaptation.

In short, our major contribution is introducing \textbf{ determinant gradient ascent (DGA), a closed-form safety index adaptation algorithm that guarantees user-defined safety for parameter-varying dynamic systems}.
For the rest of the paper, we review the literature in \Cref{sec:related}.
In \Cref{sec:pre}, we introduce the goal of safe control and the full SIS problem before formulating the problem of safety index adaptation.
In \Cref{sec:SIA}, we derive our efficient SIA approach which is then validated via a numerical study in \Cref{sec:exp}.
We finally provide future directions and conclude with \Cref{sec:conclusion}.




\section{Related Work}\label{sec:related}
Previous works \cite{ma2022joint,dawson2022safe} address SIS for known dynamics.
SIS is similar to CBF synthesis for enforcing constraints \cite{ames2014control}, but different in that the desired safety index refers to a specific class of energy functions usually for collision avoidance with the safe set algorithm (SSA) \cite{liu2014control, lin2017real}.

Real-world system dynamics are usually imperfectly known (\ie uncertainty exists). 
To address those, \cite{taylor2020adaptive} introduces adaptive CBF (aCBF) to ensure the safety of dynamic systems with estimated parametric model uncertainty. 
\cite{9129764} introduces robust aCBF (RaCBF), which results in a less conservative safe control behavior than aCBF.
\cite{10156496} applies adaptive control to CBF for safe control of systems with parametric uncertainty by adjusting the adaptation gain online. 
\cite{9196709,pmlr-v168-brunke22a,9888130} assume bounded dynamics noise and use learning-based approaches to synthesize the CBF of the mismatched system dynamics. 
\cite{9867633} focuses on high relative degree safety constraints for systems with dynamics uncertainty. It leverages concurrent learning to estimate the system uncertainty parameters online and synthesizes CBF.
\cite{https://doi.org/10.1002/rnc.6624} addresses high-order CBF for time-varying system dynamics and state constraints. 
However, these works do not consider control bounds, which are important in real-world systems and could violate safety guarantees.

\cite{zhao2021zeroviolation,zhao2023sos,wei2022nndm} address SIS for known systems with invariant bounded control.
\cite{9410332} introduces time-varying penalty functions to construct adaptive CBF when addressing systems with noisy dynamics and time-varying control bounds.
Recent work \cite{chen2023sis} addresses the SIS problem for dynamic systems with varying (\ie state-dependant) control bounds. 
Despite the rapid advancement in the field, existing works do not consider systems with both \textit{varying dynamics} and \textit{varying control bounds}, which will be addressed in this paper.

\section{Preliminaries and Problem Formulation}\label{sec:pre}


\subsection{Dynamic System}

We follow \cite{chen2023sis} and consider a dynamic system with state-dependent control limits.
Let $x\in\cX\subset \RR^{N_x}$ be the system state
and $u\in\cU$ be the control input.
The state space $\cX$ is bounded by a set of inequalities $\cX\defeq\{x\mid h_i(x) \geq 0, \forall i=1,\dots,N_h\}$.
The control space is bounded element-wise, i.e., $\cU\defeq\{u\in\RR^{N_u} \mid \underline{u} \leq u \leq \bar{u} \}$.
The dynamics is given by
\begin{equation}\label{eq:dynamics}
    \dot{x} = f(x) + g(x)u, ~ u \in \cU,
\end{equation}
where $f: \RR^{N_x} \mapsto \RR^{N_x}$ and $g: \RR^{N_x} \mapsto \RR^{N_x\times N_u}$ are both locally Lipschitz continuous.

\subsection{Preliminary: Safe Control}\label{sec:formulation_safe_control}

\textbf{Safety Specification:}
For safety, we require the state to stay within a closed subset $\cX_S$ (i.e., \textit{safe set}) of the state space $\cX$.
$\cX_S$ is assumed to be the zero sublevel set of some piecewise smooth function $\phi_0\defeq \cX \mapsto \RR$, i.e., $\cX_S \defeq \{x\in \cX \mid \phi_0(x) \leq 0\}$.
Both $\cX_S$ and $\phi_0$ should be designed by users.
For instance, $\phi_0$ can be $\phi_0 = d_\mathrm{min} - d$ if we were to keep the distance $d$ to some obstacle above $d_\textrm{min}$.

\textbf{Safe Control Objectives:}
Following \cite{chen2023sis}, we focus on safe control with two objectives: (a) {\textit{forward invariance (FI)}}, meaning if the state $x$ is already within the safe set, it should never leave that set
and (b) {\textit{finite-time convergence (FTC)}}, meaning if the state $x$ is outside the safe set, it should land in the safe set in finite time.

\textbf{Safe Control Backbone:}
When the control $u$ does not appear in $\dot\phi_0$ (e.g., $\dot\phi_0 =  - \dot d$ does not depend on the acceleration input for a second-order system), we cannot derive constraints on $u$ to ensure safety.
To solve that issue, the safe set algorithm (SSA) \cite{liu2014control} provides a systematic approach to design an alternative safety quantification $\phi$ to handle general relative degrees ($>1$) between $\phi_0$ and the control.
SSA introduces a continuous, piece-wise smooth energy function $\phi \defeq \cX \mapsto \RR$ (a.k.a. the {\textit{safety index}}).
The general form of an $n^\mathrm{th}$ ($n\geq 0$) order safety index $\phi_n$ is given as
$\phi_n = (1+a_1 s)(1+a_2 s)\dots(1+a_n s)\phi_0$ where $s$ is the differentiation operator.
$\phi_n$ is alternatively expanded to
\begin{equation}\label{def:phi_root}
    \phi_n \defeq \phi_0 + \textstyle\sum_{i=1}^{n}k_i \phi^{(i)}_0.
\end{equation}
where $\phi_0^{(i)}$ is the $i^\mathrm{th}$ time derivative of $\phi_0$.
The safe control law $c_{\phi_n}$ of SSA can be written as the following optimization:
\begin{equation}\label{eq:safe_control_law}
    \minimizewrt{u\in\cU} \cJ(u) ~ \st ~ \dot{\phi}_n(x,u) \leq -\eta~\mathrm{if}~\phi_n(x) \geq 0 
\end{equation}
where the objective $\cJ$ is arbitrary.
By \cite{liu2014control,chen2023sis}, if (a) the roots of the characteristic equation $\prod_{i=1}^n(1+a_i s) = 0$ are all negative real,
(b) $\phi_0^{(n)}$ has relative degree one to the control input, and (c) the problem \eqref{eq:safe_control_law} is always feasible, both FI and FTC are guaranteed.
\revcolor{Note that \eqref{eq:safe_control_law} only considers constraint satisfaction which is compatible with arbitrary control objectives.
For instance, for reference tracking, we can set $\cJ(u) = \|u-u^r\|$ to find $u$ that is minimally invasive to the nominal control $u^r$, presumably generated by a given tracking controller with asymptotical stability.}


\subsection{Preliminary: Safe Index Synthesis}\label{sec:SIS_formulation}

To achieve safety guarantees by implementing \eqref{eq:safe_control_law}, we need to construct $\phi$ to make the optimization feasible.
Such an objective is referred to as \textit{Safety Index Synthesis} (SIS), mathematically described as \Cref{problem:synthesis}.
\begin{problem} [Safety Index Synthesis] \label{problem:synthesis}
    Find safety index as $\phi_\theta \defeq \phi_0 + \sum_{i=1}^{n}k_i \phi^{(i)}_0$ with parameter $\theta\in\Theta\defeq\{[k_1,k_2,\dots,k_{n}] \mid k_i\in\RR,k_i\geq 0,\forall i \}$, such that
    \begin{equation}\label{eq:synthesis}
        \forall x\in\cX ~ \st ~ \phi_\theta(x)\geq 0, \minimizewrt{u\in\cU} \dot{\phi}_\theta(x,u) < -\eta.
    \end{equation}
\end{problem}
$\phi_\theta$ is the $n^\mathrm{th}$ order safety index parameterized by $\theta$ and is used interchangeably with $\phi_n$ hereafter for clarity.
Note that \Cref{problem:synthesis} depends on the dynamics \eqref{eq:dynamics} (i.e., $f$ and $g$) since $\dot{\phi}_\theta(x,u) = \frac{\partial\phi_\theta}{\partial x}f(x) + \frac{\partial\phi_\theta}{\partial x}g(x) u$ in \eqref{eq:synthesis}.
\Cref{problem:synthesis} is also difficult for having infinitely many constraints since \eqref{eq:synthesis} needs to hold for \textbf{any} state $x~\st \phi_\theta(x) \geq 0$.
To tackle that challenge, we follow \cite{zhao2023sos,chen2023sis} and leverage Positivstellensatz \cite{parrilo2003sosp} to transform \Cref{problem:synthesis} into a sum-of-square programming (SOSP) 
which is further converted to nonlinear programming (NP).
In specific, a refute set $\{x\mid \zeta_{i=1,\dots,N_\zeta}(x)=0,\gamma_{i=1,\dots,N}(x)\geq 0\}$ is first established for \eqref{eq:synthesis}, then proved empty by solving an SOSP.
We refer readers to \cite{chen2023sis} for details on the construction of the refute set.
The SOSP finds $p'_i \in \RR, p_i \geq 0 ,~ \forall i>0$ such that
\begin{equation}\label{eq:nonlinear_sos}
    \begin{aligned}
        &p_0 = -1 - \textstyle\sum_{i=1}^{N_\zeta}p'_i\zeta_i - p_1\gamma_1 - p_2\gamma_2 - \dots - p_N\gamma_N \\
        &~~~~~~~~ -p_{12}\gamma_{1}\gamma_{2} - \dots - p_{12\dots N}\gamma_{1}\dots\gamma_{N} \in SOS.
    \end{aligned}
\end{equation}
where $\zeta_i$, $\gamma_i$ are functions of $x$ and also depend on $f$ and $g$.
The SOS condition is enforced by finding the positive semi-definite (PSD) decomposition $p_0 = \bx^\top Q(\theta,\bp)\bx$ where $Q(\theta,\bp)\succeq 0$.
Assuming $p_0$ has degree $2d$, $\bx\defeq\left[1,x[1],\dots,x[N_x],x[1]x[2],\dots,x[N_x]^d\right]^\top$ contains all monomials of $x$ with order no more than $d$.
$\bp\defeq [p'_1,\dots,p'_{N_\zeta}, p_1,p_2,\dots,p_{012\dots N}]^\top$ denotes the auxiliary decision variable.
The final NP is given by:
\begin{problem}[Nonlinear Programming]\label{problem:nonlinear}
    Find $\theta\in\Theta$ and $\bp$ where $\bp[j] \in \RR$ for $j > 0$ and $\bp[j] \geq 0$ for $j > N_\zeta$, such that $Q(\theta, \bp) \succeq 0$.
\end{problem}

\begin{remark}\label{rmk:dyn_dep}
The positive-semidefiniteness of the parametric coefficient matrix $Q(\theta, \bp)$ guarantees the positiveness of polynomial $p_0$.
Since $Q$ is derived from $\zeta_i$ and $\gamma_i$, it depends on the dynamics \eqref{eq:dynamics}, i.e., $f$, $g$, and the control limits $\cU$.
\end{remark}

\begin{remark}
The general form of the SOSP \eqref{eq:nonlinear_sos} allows $p_i'$ and $p_i$ to be polynomials of $x$.
Hence, due to the simplifications (\ie constraining $p_i'$ and $p_i$ to real values), \Cref{problem:nonlinear} solves a sufficient but not necessary condition to \Cref{problem:synthesis}.
\end{remark}

\subsection{Formulation of Safety Index Adaptation}


As motivated in \Cref{sec:intro}, practical dynamic systems can contain varying parameters only known during runtime.
We denote varying parameters as $\rho$ and extend \eqref{eq:dynamics} as
\begin{equation}\label{eq:dynamics_param}
    \dot{x} = f(x,\rho) + g(x,\rho)u, ~ u \in \cU(\rho).
\end{equation}

Assume that prior to deployment, the initial value $\rho_0$ is known, and a feasible safety index $\phi_{\theta_0}$ has been solved via \Cref{problem:synthesis}.
As explained in \Cref{rmk:dyn_dep}, $\phi_\theta$ depends on the system dynamics.
With the extended dynamics \eqref{eq:dynamics_param}, $\phi_\theta$ also depends on $\rho$.
As a result, when $\rho$ is updated during runtime, the previously solved $\phi_\theta$ might no longer satisfy the feasibility condition \eqref{eq:synthesis} and render the system unsafe.
Hence, it is imperative that $\phi_\theta$ is updated accordingly, formulated as:
\begin{problem}[Safety Index Adaptation (SIA)]\label{problem:SIA}
    Given a solution $\phi_{\theta}$ to \Cref{problem:synthesis} with system parameter $\rho$, find $\phi_{\theta'}$ to solve \Cref{problem:synthesis} with system parameter $\rho'$\footnote{We assume bounded step changes in the system parameters, \ie $\|\rho-\rho'\| \leq \delta$ for some $\delta>0$. Theoretical results on how the step size $\delta$ influences the adaptation performance are left for future work.}.
\end{problem}

\begin{remark}
    A naive solution to \Cref{problem:SIA} is to directly re-run the full synthesis given by \Cref{problem:nonlinear}.
    However, the solving time of the NP is significant even for simplistic systems, e.g., over $10$ minutes for a second-order unicycle model \cite{chen2023sis}.
    For safety-critical tasks, the safety guarantees of the safe control law should be recovered as soon as possible.
\end{remark}


\section{SIA via Determinant Gradient Ascend}\label{sec:SIA}


Although re-running the full synthesis (\Cref{problem:nonlinear}) is infeasible, we can leverage the NP formulation to design an adaptation strategy.
Observe that solving \Cref{problem:synthesis} is ultimately achieved by making the parametric coefficient matrix $Q(\theta, \bp, \rho)\succeq 0$, where the dependency on $\rho$ follows \eqref{eq:dynamics_param}.
Then, \Cref{problem:SIA} naturally translates to:
\begin{equation}\label{optim:SIA}
    \begin{aligned}
        \theta',\bp' = \argminwrt{\theta,\bp} ~ \cJ(\theta,\bp) ~ \st Q(\theta,\bp,\rho') \succeq 0
    \end{aligned}
\end{equation}
given $Q(\theta,\bp,\rho) \succeq 0$, where the objective $\cJ$ is a design parameter to guide the search for $\theta$ and $\bp$.
If $\rho'$ does not change significantly from $\rho$, i.e., $\|\rho-\rho'\|$ is bounded (to be formalized later), we are essentially searching for a new point $(\theta',\bp',\rho')$ near the neighborhood of $(\theta,\bp,\rho)$ to maintain the positive-semidefiniteness of $Q$.

Note that the positive-semidefiniteness of $Q$ can be tested by computing determinants using Sylvester's criterion \cite{horn2012matrix}, which says that a Hermitian matrix is positive-semidefinite if and only if all the principal minors are nonnegative.
Namely, we can re-write the constraint in \eqref{optim:SIA} as
\begin{equation}\label{optim:SIA_det_constraint}
\begin{aligned}
    \mathrm{Det}[Q(\theta,\bp,\rho')]_{I,I} \geq 0,~\forall I\subseteq[1,\dots,M]
\end{aligned}
\end{equation}
where $M$ is the size of $Q$.
$[Q]_{I,J}$ denotes the submatrix of $Q$ corresponding to the rows with indices $I$ and columns with indices $J$.
Since the principal minors are essentially explicit functions of $\theta$ and $\bp$, \eqref{optim:SIA_det_constraint} can be readily satisfied via gradient ascends on those parameters as $[{\theta},{\bp}]=[{\theta},{\bp}]+\lambda\delta$ where the gradient $\delta$ is given by
\begin{equation}\label{eq:SIA_grad}
    \begin{aligned}
        \delta = \left.\nabla_{[\theta,\bp]} \mathrm{Det}[Q(\theta,\bp,\rho')]_{I^*,I^*}\right\rvert_{\theta=\theta,\bp=\bp}.
    \end{aligned}
\end{equation}
$\mathrm{Det}[Q(\theta,\bp,\rho')]_{I^*,I^*}$ refers to the current lowest principal minor with indices $I^*$ and $\lambda$ the step size.
Upon change of $\rho$, $(\theta',\bp')$ is initialized to the previous feasible values $(\theta,\bp)$, and updated according to \eqref{eq:SIA_grad} until all principal minors are nonnegative.
We refer to such an approach as the determinant gradient ascend (DGA).
\revcolor{\textit{After} $\phi_n$ is fully updated for $\rho'$, \eqref{eq:safe_control_law} would be feasible and guarantee FI and FTC with respect to $\cX_S$.
Future work remains to study the system behaviors \textit{during} DGA adaptation, when \eqref{eq:safe_control_law} might be infeasible.}

\begin{remark*}
\revcolor{Since $Q$ depends on $f$ and $g$ which are fixed functions of $x$ and $\rho$, the form of gradient update \eqref{eq:SIA_grad} is also fixed.}
Hence, with a pre-computed symbolic expression of the update, one only has to evaluate \eqref{eq:SIA_grad} on different $(\theta,\bp,\rho)$ values during deployment, which is fast enough to support real-time adaptation.
In summary, the determinant gradient ascend (DGA) method enables close-form solutions to safety index adaptation using previous indices for warm start.
\end{remark*}




\section{Numerical Study}\label{sec:exp}
To validate our SIA approach, we provide a numerical study on a parameter-varying system based on a 2-DOF (degree of freedom) planar robot arm.
The robot arm has a second-order dynamics model with joint acceleration as the input.
We first derive the baseline NP problem for SIS following \Cref{sec:SIS_formulation} and then derive the update rule for SIA in the form of \eqref{eq:SIA_grad}.
The feasibility of the adaptive safety index is validated by sample based evaluations.


\subsection{Parameter-varying 2-DOF Robot Arm}

\begin{wrapfigure}{r}{0.5\linewidth}
    \vspace{-5pt}
    \centering
    \includegraphics[width=\linewidth]{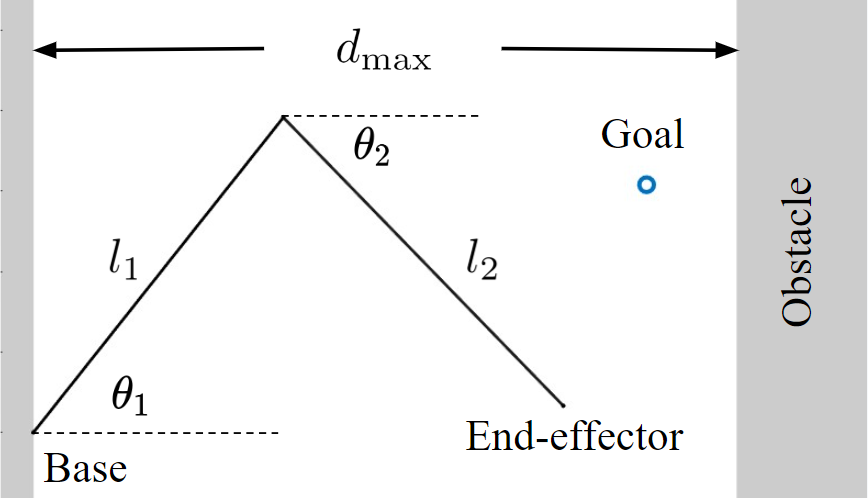}
    \caption{2-DOF Robot Arm.}
    \vspace{-10pt}
    \label{fig:2d_arm}
\end{wrapfigure}
We consider a 2-DOF robot arm with state $x\defeq[\theta_1, \theta_2, \dot{\theta}_1, \dot{\theta}_2]^\top$, where $\theta_{1,2}\in[-\pi/2, -\pi/18] \cup [\pi/18, \pi/2]$ are the joint positions as shown in \Cref{fig:2d_arm}.
$\dot{\theta}_{1,2}\in[-1,1]$ are joint velocities.
The two links have length $l_1$ and $l_2$ respectively.
The control $u\defeq[u_1,u_2]^\top$ includes bounded joint acceleration input $u_{1,2}\equiv \ddot{\theta}_{1,2} \in[u_\mathrm{min},u_\mathrm{max}]$.
The dynamics of the 2-DOF robot is given by $\dot{x}=f(x)+g(x)u$ where
\begin{equation}\label{eq:exp_dynamics}
    f(x) = \begin{bmatrix}
        \dot{\theta}_1 \\ \dot{\theta_2} \\ 0 \\ 0
    \end{bmatrix},~g(x) = \begin{bmatrix}
        0 & 0 \\ 0 & 0 \\ 1 & 0 \\ 0 & 1
    \end{bmatrix}
\end{equation}
In real-world scenarios, system dynamics might change due to external factors.
For instance, the total mass of a drone changes with different payloads, which in turn changes its dynamics; the torque limit of an arm motor might change due to insufficient power supply.
In those cases, safety index adaptation is necessary to guarantee safety.
Hence, to verify our SIA approach, we extend \eqref{eq:exp_dynamics} to an affine parameter-varying system
\begin{equation}\label{eq:exp_dynamics_param_vary}
    \dot{x} = f(x,\rho) + g(x,\rho)u = A^f f(x) + A^g g(x) u + b
\end{equation}
where $A^f=\bI$, $A^g=\mathrm{diag}([1, 1, c_1, c_2])$ and $b=[0, 0, b_1, b_2]^\top$.
We assume $c_{1,2} \geq 0$ and $b_{1,2} \in \RR$.
The parameters $\rho\defeq[c_{1,2}, b_{1,2}]$ are the system parameters, which can change during runtime and can be directly observed.
The robot is allowed to move within the free space and should not collide with the obstacle which is a wall placed $d_\mathrm{max}$ from the robot base.

\subsection{Safety Index Adaptation Rule}


We first derive the full SIS solution which is required to derive DGA update rules.
With $\phi_0 = l_1\cos(\theta_1)+l_2\cos(\theta_2)-d_\mathrm{max}$, SIS produces a safety index $\phi_\theta = \phi_0 + k \dot{\phi}_0$ such that the control law \eqref{eq:safe_control_law} always keeps the end-effector at most $d_\mathrm{max}$ away horizontally from the base, not colliding with the wall.
The SI parameter $\theta$ contains a single parameter $k\geq 0$.
The immediate next step is to write out the feasibility condition \eqref{eq:synthesis} to be met by $\phi_\theta$.
We first handle the main condition $\mathbf{min}_{u\in\cU} \dot{\phi}_\theta(x,u) < -\eta$.
Plugging in $\phi_0$, we have
\begin{equation*}
    \phi_\theta = l_1\cos\theta_1 + l_2\cos\theta_2 - k l_1 \sin\theta_1 \dot{\theta}_1 - k l_2 \sin\theta_2 \dot{\theta}_2 - d_\mathrm{max}.
\end{equation*}
Taking time derivative, we have
\begin{equation*}
    \begin{aligned}
        \dot{\phi}_\theta =& \sum_{j=1,2} -l_j \sin\theta_j \dot{\theta}_j - k l_j\cos\theta_j \dot{\theta}_j^2 - k l_j\sin\theta_j \ddot{\theta}_j \\
        =& \textstyle\sum_{i=1,2} -l_j \sin\theta_j \dot{\theta}_j - k l_j\cos\theta_j \dot{\theta}_j^2 - k l_j\sin\theta_j(c_j u_j + b_j)
    \end{aligned}
\end{equation*}
Note that $k,l_j,c_j\geq0$, hence the minimum of $\dot{\phi}_\theta$ is reached at $u_j = u_\mathrm{max}$ if $\sin\theta_j \geq 0$ and $u_j = u_\mathrm{min}$ otherwise.
\revcolor{Since $\theta_j\in[-\pi/2, -\pi/18] \cup [\pi/18, \pi/2]$, the positiveness of $\theta_j$ depends on which interval it falls into, namely whether $\theta_j \leq -\pi/18$ or $\theta_j \geq \pi/18$.
With indicators $\II_{1,2}=\pm 1$, those conditions can be written as}
\begin{align}
    \II_{j}\sin\theta_j - \sin(\pi/18) \geq 0 \label{eq:gamma_min_angle}
\end{align}
Then, the main feasibility condition becomes
\begin{align}
    &\textstyle\sum_{j=1,2} -l_j \sin\theta_j \dot{\theta}_j - k l_j\cos\theta_j \dot{\theta}_j^2 \nonumber \\
    &~~~~~~~~~ - k l_j\sin\theta_j(c_j \tilde{u}_j + b_j) < -\eta \label{eq:gamma1}
\end{align}
where $\tilde{u}_j=u_\mathrm{max}$ if $\II_{j}=1$ and $\tilde{u}_j=u_\mathrm{min}$ if $\II_{j}=-1$ for $j=1,2$.
Next, we add conditions to consider the state limits, \ie $\theta_j\in[\pi/18,\pi/2]$, $\dot{\theta}_{j} \in [-1, 1]$ and $\dot{\theta}_{j}^2 \in [0, 1]$:
\begin{align}
    -\II_j \sin\theta_j + 1 &\geq 0 \label{eq:gamma_sin_bound}\\
    1 - \dot{\theta}_j^2 &\geq 0 \label{eq:gamma_vel_bound}\\
    -(\dot{\theta}_j^2)^2 + \dot{\theta}_j^2 &\geq 0 \label{eq:gamma_vel2_bound}\\
    \sin\theta_j^2 + \cos\theta_j^2 - 1 &= 0 \label{eq:gamma_sincos}
\end{align}
The last condition in \eqref{eq:synthesis} is $\phi_\theta \geq 0$, which is omitted here to enable decreasing safety index at all levels (\ie $\phi_\theta \in \RR$), instead of only the unsafe regions (\ie $\phi_\theta \geq 0$).
Now, \eqref{eq:synthesis} translates to: for any state satisfying \eqref{eq:gamma_min_angle} to \eqref{eq:gamma_sincos}, \eqref{eq:gamma1} holds.
To achieve that, we construct a refute set by collecting \eqref{eq:gamma1} to \eqref{eq:gamma_sincos}, with \eqref{eq:gamma1} negated, and prove that the refute set is empty\footnote{See \cite{chen2023sis} for the theoretical results of such an approach.}.
With $\alpha_{j}\defeq \sin\theta_j$, $\beta_{j}\defeq \cos\theta_j$, $y_j\defeq \dot{\theta}_j$ and $z_j\defeq\dot{\theta}_j^2$ for $j=1,2$, the refute set is given by:
\begin{equation}\label{eq:example_refute}
    \begin{cases}
        \gamma_1\defeq -\l_1 \alpha_1 y_1 - k l_1 \beta_1 z_1 - k l_1 (c_1\tilde{u}_1+b_1)\alpha_1 \\
        ~~~~ -\l_2 \alpha_2 y_2 - k l_2 \beta_2 z_2 - k l_2 (c_2\tilde{u}_2+b_2)\alpha_2 \geq 0 \\
        \gamma_2\defeq \II_1\alpha_1 - \sin(\pi/18) \geq 0 \\
        \gamma_3\defeq -\II_1\alpha_1 + 1\geq 0\\
        \gamma_4\defeq 1-y_1^2 \geq 0\\
        \gamma_5\defeq -z_1^2+z_1\geq 0\\
        \zeta_1\defeq \alpha_1^2+\beta_1^2-1=0\\
        \gamma_6\defeq \II_2\alpha_2 - \sin(\pi/18) \geq 0 \\
        \gamma_7\defeq -\II_2\alpha_2 + 1\geq 0\\
        \gamma_8\defeq 1-y_2^2 \geq 0\\
        \gamma_9\defeq -z_2^2+z_2\geq 0\\
        \zeta_2\defeq \alpha_2^2+\beta_2^2-1=0
    \end{cases}
\end{equation}
The refute set is represented by four versions of \eqref{eq:example_refute} with different sign values of $\II_{1,2}$.
Following \eqref{eq:nonlinear_sos}, for the $i^\mathrm{th}$ assignment ($i\in[4]$) of $(\II_{1}, \II_{2})$, we have
\begin{equation}
\begin{aligned}
    p_{i,0} = -1-p'_{i,1}\zeta_{i,1}-p'_{i,2}\zeta_{i,2}-\textstyle\sum_{n=1}^{9} p_{i,n}\gamma_{i,n}
\end{aligned}
\end{equation}
and decompose as $p_{i,0}= \bx^\top Q_i(\theta,\bp_i,\rho)\bx$ where $\bx\defeq[1, y_1, z_1, \alpha_1, \beta_1, y_2, z_2, \alpha_2, \beta_2]^\top$, $\theta\defeq[k]$, and $\bp_i\defeq [p'_{i,1}, p'_{i,2}, p_{i,1},\dots,p_{i,9}]$.
Let $[Q]_{m,n}$ denote the element of $Q$ at row $m$ column $n$, we have
\begin{equation}\label{eq:exp_Q}
    \begin{cases}
        [Q_i]_{2,4} = -l_1 p_{i,1} \\
        [Q_i]_{3,5} = -k l_1 p_{i,1} \\
        [Q_i]_{1,4} = -k l_1 (c_1\tilde{u}_{i,1}+b_1)p_{i,1} + \II_{i,1} p_{i,2} - \II_{i,1} p_{i,3} \\
        [Q_i]_{4,4} = p'_{i,1} \\
        [Q_i]_{2,2} = -p_{i,4} \\
        [Q_i]_{3,3} = -p_{i,5} \\
        [Q_i]_{1,3} = p_{i,5} \\
        [Q_i]_{5,5} = p'_{i,1} \\
        [Q_i]_{6,8} = -l_2 p_{i,1} \\
        [Q_i]_{7,9} = -k l_2 p_{i,1} \\
        [Q_i]_{1,8} = -k l_2 (c_2\tilde{u}_{i,2}+b_2)p_{i,1} + \II_{i,2} p_{i,6} - \II_{i,2} p_{i,7} \\
        [Q_i]_{8,8} = p'_{i,2} \\
        [Q_i]_{6,6} = -p_{i,8} \\
        [Q_i]_{7,7} = -p_{i,9} \\
        [Q_i]_{1,7} = p_{i,9} \\
        [Q_i]_{9,9} = p'_{i,2} \\
    \end{cases}
\end{equation}
With that in hand, the gradient updates \eqref{eq:SIA_grad} can be obtained by taking derivatives of the principal minors of $\{Q_i\}_{i=1,\dots,4}$ with respect to $[\theta,\bp_1,\dots,\bp_4]$.
Specifically, given new parameter $\rho'$ to adapt to, we compute the gradients as:
\begin{equation}
    \begin{aligned}
        \delta_\theta &= \frac{1}{4} \sum_{i=1}^4 \left.\nabla_{\theta} \mathrm{Det}[Q_i(\theta,\bp_i,\rho')]_{I^*_i,I^*_i}\right\rvert_{\theta=\theta,\bp_i=\bp_i} \\
        \delta_{\bp_i} &= \left.\nabla_{\bp_i} \mathrm{Det}[Q_i(\theta,\bp_i,\rho')]_{I^*_i,I^*_i}\right\rvert_{\theta=\theta,\bp_i=\bp_i}
    \end{aligned}
\end{equation}
With learning rate $\lambda_k$ and $\lambda_{\bp}$, we apply the update rule
\begin{equation}\label{eq:exp_update_rule}
    \begin{aligned}
        \theta = \theta + \lambda_\theta \delta_\theta, ~\bp_i = \bp_i + \lambda_{\bp} \delta_{\bp_i}
    \end{aligned}
\end{equation}
until all principal minors of all $Q_i$'s are non-negative.


\begin{figure}[t]
    \centering
    \begin{subfigure}[b]{0.45\linewidth}
        \centering
        \includegraphics[width=\linewidth]{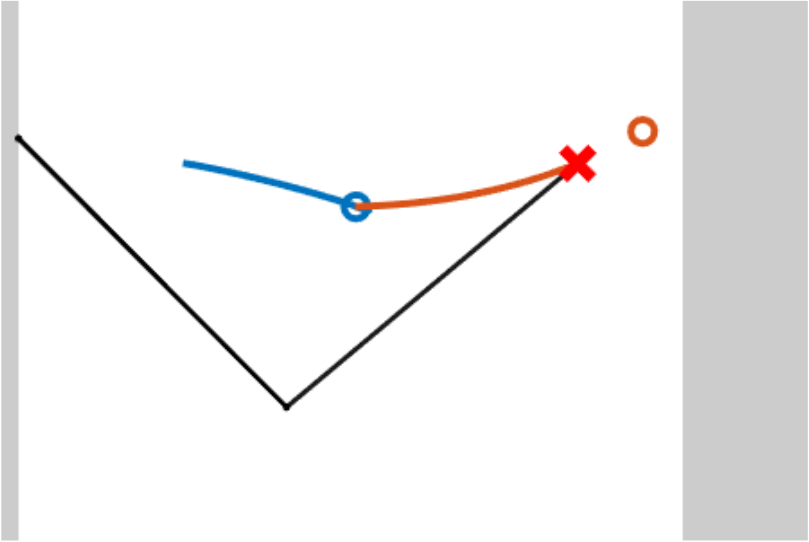}
        \caption{Without adaptation.}
    \label{fig:demo_no_adapt}
    \end{subfigure}
    \hfill
    \begin{subfigure}[b]{0.45\linewidth}
        \centering
        \includegraphics[width=\linewidth]{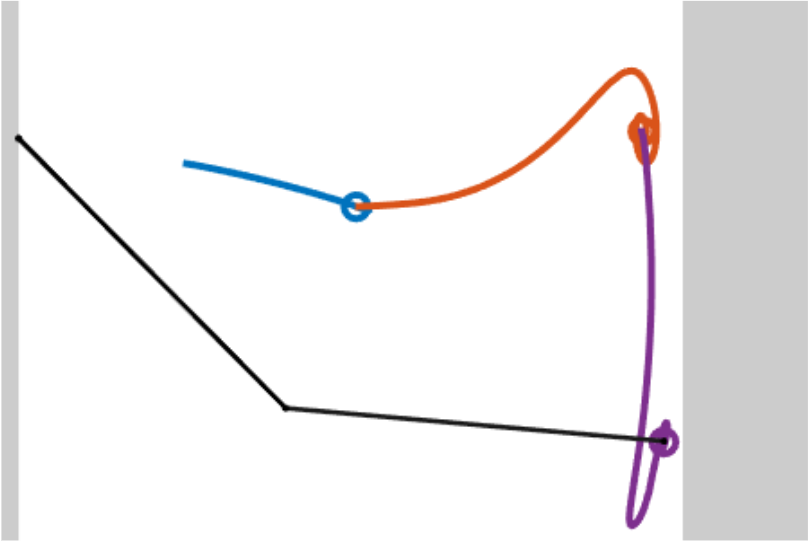}
        \caption{With adaptation}
    \label{fig:demo_adapt}
    \end{subfigure}
    \caption{Arm end-effector tracking without and with safety index adaptation.
    Each goal is marked with the same color as the corresponding tracking trajectory.
    The robot is initialized with a feasible safety index with respect to the initial system dynamics and starts to track the first goal in blue.
    Every time a goal is reached, the system dynamics change.
    (a) Without adaptation, when tracking the second goal in orange, the arm runs into a state (marked by a cross) where it is approaching the wall quickly and no safe control can be found within the control limits.
    (b) With adaptation, the safety index is updated upon changes to the dynamics.
    That keeps the safe control law always feasible.
    As a result, the arm decelerates in advance when approaching the wall and safely tracks each of the goals.
    }
    \label{fig:demo}
    \vspace{-10pt}
\end{figure}

\begin{figure}[t]\label{fig:test_result}
    \centering
    \includegraphics[width=\linewidth]{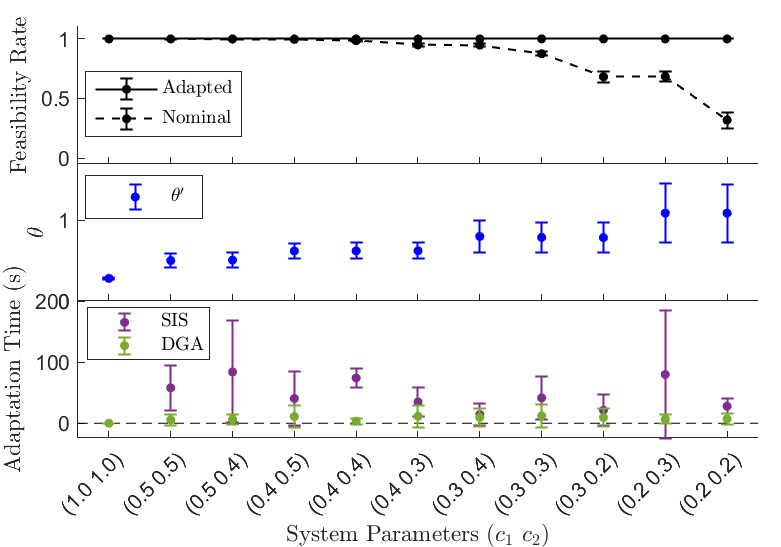}
    \caption{Feasibility rate of adapted safety indices, safety index parameters $\theta'=[k']$ and adaptation time under different system parameters $\rho'$. The first point ($\rho: c_1=c_2=1.0$) corresponds to the nominal system. $b_1$ and $b_2$ are always $0$.}
    \label{fig:feas_rate}
    \vspace{-10pt}
\end{figure}

\subsection{Experiment and Results}

We initialize the robot arm with nominal parameters $\rho = [c_1=c_2=1, b_1=b_2=0]$ and run the full safety index synthesis (see \Cref{problem:nonlinear}) to acquire an initial safety index $\phi_\theta$.
The inputs are limited to $u_\mathrm{min}=-100,u_\mathrm{max}=100$.
To validate our SIA approach, we simulate multiple disturbances to the system parameters $\rho$.
For each perturbed system with parameters $\rho'$, we invoke the SIA update rules \eqref{eq:exp_update_rule} to acquire a new $\phi_{\theta'}$.
\Cref{fig:demo} shows an example of such adaptation where the parameters $\rho$ is perturbed after the arm end-effector reaches its goal.
Without adaptation, the arm runs into a state where the safe control law \eqref{eq:safe_control_law} is infeasible and fails to ensure safety.
With adaptation, the arm quickly updates the SI to $\phi_{\theta'}$ and manages to find safe actions.

For quantitative evaluation, we apply each adapted $\phi_{\theta'}$ by running the safe control law \eqref{eq:safe_control_law} on $1000$ uniformly sampled states under the perturbed system and compare to the nominal safety index $\phi_\theta$.
If \eqref{eq:safe_control_law} is feasible, we mark the safety index as feasible at the corresponding state.
Due to the uncertainty of nonlinear programming \Cref{problem:nonlinear}, we repeat the whole process for $10$ times and plot the feasibility rate of the safety index before and after adaptation, the adapted SI parameter $\theta$ and adaptation time.
We also run the full SIS on each perturbed system and compare the computation time.
See \Cref{fig:feas_rate} for the plots.
We observe that the more $\rho'$ deviates from $\rho$ (the smaller the $c_{1,2}$), the control law under the nominal safety index is less likely to be feasible while the adapted safety index achieves $100\%$ feasibility rate.
The adaptation time is also consistently lower than that of solving full SIS, validating that our SIA approach is computationally efficient for real-time deployment.
\revcolor{Although only $c_{1,2}$ are perturbed in our simulations, our approach directly accommodates other variations, for instance changing $b_{1,2}$ or more generally, changing $A^f$, $A^g$ and $b$ in \eqref{eq:exp_dynamics_param_vary}.}

\subsection{Discussions}


\textbf{Tolerance against variations.}
It can be observed from \Cref{fig:test_result} that the adapted value of SI parameter $k$ shows a negative correlation with respect to the system parameters $c_{1,2}$.
In our experiments, we discovered that when $c_{1,2}$ are increased, the original $k$ is normally still feasible, and no adaptation is required.
Intuitively, the larger $c_{1,2}$ is, the more sensitive the system is to inputs; the larger $k$ is, the more sensitive the control law is to unsafe regions.
When $c_{1,2}$ increases, the system becomes more reactive, keeping the original $k$ feasible.
When $c_{1,2}$ decreases, a more aggressive safe control law is needed to react to unsafe regions in advance, necessitating a larger $k$.
Note that the above only applies to our specific system, while the tolerance analysis for general systems is left for future work.

\textbf{Scalability against system dimensions.}
The scalability of both full SIS and SI adaptation largely depends on the size of the refute set \eqref{eq:example_refute} as well as the coefficient matrix $Q_i$ in \eqref{eq:exp_Q}.
For an $n$-DOF 2D robot arm, the size of the refute set is given by $1+5n$; the size of $Q_i$ is $1+4n$; and there are $2^n$ such $Q_i$ to prove PSD for full SIS.
Despite the exponential scalability of SIS, our DGA approach allows one to pre-generate all gradient updates from $Q_i$ in symbolic forms and only evaluates those expressions during online adaptation.
That renders our approach highly efficient even for high-dimensional systems.

\textbf{Gradient-based Optimization.}
When implementing our update rule \eqref{eq:exp_update_rule}, we normalize the gradients $\delta_\theta$ and $\delta_{\bp_i}$, and set the learning rates $\lambda_\theta=\lambda_{\bp}=1e-5$.
Empirically, one should always normalize the gradients and start experimenting with small learning rates to help DGA converge.
Moreover, our DGA is presented in first-order gradient updates in \eqref{eq:exp_update_rule}.
Second-order approaches such as Newton's method can also be applied for better convergence rates when the change of $\rho$ is minimal and a feasible $k'$ can be found within a near neighbor of the current $k$.

\section{Conclusion and Future Work}\label{sec:conclusion}

In this paper, we presented a safety index adaptation (SIA) approach to update safe control laws in response to varying system dynamics in real time.
Our approach replaces full safety index synthesis, which is extremely slow, with fast closed-form updates to controller parameters.
Through numerical studies, we verified that our approach allows the agent to quickly adapt to new system dynamics and achieve zero safety violations.

In practice, after the system dynamics change, the system is inevitably guarded by an outdated safety index during the adaptation computation time.
Hence, as future work, it is worth studying the system's behavior during such a transition period to draw critical insights, for instance, whether the adaptation can finish before the agent crashes into unsafe regions.
If not, the agent should stop navigation and wait for the new safety index.
Another promising direction is to handle continuously changing dynamics as opposed to step parameter changes, which will bring new questions on the tolerance of synthesized safety indices and the criterion of triggering SIA.
Finally, we aim to provide theoretical results such as the proof of convergence to the new safety index as well as the convergence rate.

\addtolength{\textheight}{-12cm}   








\bibliographystyle{IEEEtran}
\bibliography{ref}

\end{document}